\documentclass[conference]{IEEEtran}
\usepackage{cite}
\usepackage{amsmath,amssymb,amsfonts}
\usepackage{graphicx}
\usepackage{textcomp}
\usepackage{xcolor}
\usepackage{listings}
\usepackage{url}

\lstdefinestyle{rust-example}{
  basicstyle=\footnotesize\ttfamily,
  columns=fullflexible,
  keepspaces=true,
  showstringspaces=false,
  numbers=left,
  numberstyle=\tiny\color{black!60},
  numbersep=5pt,
  xleftmargin=1.25em,
  morekeywords={fn,self,Self,usize},
  keywordstyle=\color{blue!65!black}\bfseries,
  morecomment=[l]{//},
  commentstyle=\color{black!65}
}   
\def\BibTeX{{\rm B\kern-.05em{\sc i\kern-.025em b}\kern-.08em
    T\kern-.1667em\lower.7ex\hbox{E}\kern-.125emX}}
\usepackage{tikz}
\usetikzlibrary{shapes.geometric}
\usetikzlibrary{shapes.multipart,positioning,arrows.meta}
\usetikzlibrary{arrows.meta,positioning}
\begin{document}

\title{Multi-Agent Coordinated Method-Level Refactoring of Rust Code}

\author{
	\IEEEauthorblockN{Yijian Zheng, Yi Zhang, Juyuan Zhang, Di Cui*\thanks{*Corresponding author.}}
	\IEEEauthorblockA{School of Computer Science and Technology, Xidian University, Xi’an 710071, China} 
	\IEEEauthorblockA{cuidi@xidian.edu.cn; 25031212388@stu.xidian.edu.cn}
}

\maketitle

\begin{abstract}
As Rust codebases evolve, developers refactor code to reduce
duplication, simplify complex methods, and keep related
functionality together. Developers can make these changes
manually or use IDE assists for supported operations, then
check the resulting code with compiler diagnostics and tests.
Method-level refactoring in Rust must preserve observable behavior
while accounting for ownership, borrowing, and lifetimes.
We present a large language model (LLM)-based workflow for
extracting, inlining, and moving methods in Rust. The workflow
combines candidate generation, configurable specialist review,
and container-based validation. On six tasks
from three open-source Rust projects, three end-to-end
configurations passed all task checks on their first attempt,
using 6, 13, and 31 model calls. To isolate the effect of review,
we then gave four configurations the same six reference patches
and six deliberately behavior-changing variants. A designated
specialist rejected all six behavior-changing patches before
container validation and let all six reference patches proceed,
reducing Docker invocations from 24 to 12 relative to direct
validation. The fixed specialist panel also caught all six faults
early and rejected two reference patches over unmeasured performance
concerns. On this workload, specialist review reduced execution
work, while its effect on total time depended on review overhead
and which patches reached validation.
\end{abstract}


\section{Introduction}
\label{sec:introduction}
As Rust libraries evolve, routines that once served a single
purpose can accumulate responsibilities, and related logic can
spread across multiple call sites. For example, a method that
originally traversed a collection may later handle input
validation and result construction as well. A maintainer changing
one of these responsibilities must then trace the surrounding
logic to determine which code the change affects. Refactoring
can make these responsibilities explicit and make subsequent
changes easier to localize~\cite{fowler2018refactoring}.
Method-level refactoring reorganizes code while preserving its
observable behavior.Extracting a code fragment moves it into a new helper function
and replaces the fragment with a call to that function. In Rust,
if the helper returns a borrowed reference, its signature may
need to specify which input the reference borrows from. Inlining
replaces a function or method call with the computation it invokes.
The rewrite must preserve how its arguments are evaluated and
when its temporary values are dropped. Moving a method across modules or crates
can change which items it can access and how callers reach it.
A tool must account for these obligations even when the developer
specifies only the intended structural change.

Prior work has addressed several of these challenges for method
extraction. REM uses compiler information to repair control-flow,
ownership, and lifetime issues introduced by
extraction~\cite{thy2023adventure}. REM-V and REM2.0 examine how
machine-checked equivalence can be established for supported
extractions~\cite{britton2025verifying,britton2026rem2}.
Multi-agent systems have also been used to plan, apply, and
validate refactorings in Java~\cite{oueslati2026refagent}.
When configurations generate different patches, differences in
their outcomes can arise from both the proposals and their review.
This study asks what specialist review contributes when the
candidate patch and downstream validation procedure are held fixed.

In this paper, we implemented a LLM-based workflow for method extraction,
inlining, and moving in Rust. The workflow sends a generated or
supplied candidate to container-based validation, with optional
specialist review beforehand. We compare direct validation, one
designated specialist, adaptive specialist selection, and a fixed
specialist panel. Our primary contribution is a paired evaluation
that presents identical candidate patches to all four configurations
and measures when faulty patches are rejected, whether reference
patches are rejected during review, and how much work each
configuration performs.

We complemented the paired evaluation with an end-to-end experiment
on six tasks from three open-source Rust projects. All 18
generation runs pass the task checks on their first attempt,
with different model-call costs across configurations. In the
48-run paired evaluation, a designated specialist rejects all six
behavior-changing patches before container validation and lets
all six reference patches proceed. The fixed panel also rejects
all six faults early and rejects two reference patches over
unmeasured performance concerns. These outcomes quantify the
trade-off between earlier rejection, review overhead, and the
risk of discarding patches that satisfy the task checks.
\section{Background and Related Work}
\label{sec:background}

\subsection{Method-Level Refactoring in Rust}

We study three operations from Fowler's refactoring
catalog~\cite{fowler2018refactoring}. We use the terms
\emph{Extract Method}, \emph{Inline Method}, and \emph{Move Method}
for the corresponding operations in our Rust tasks.Chen et al.~\cite{chen2025panguembedded} introduce Pangu Embedded,
a reasoning model that supports fast and slow inference modes
to balance response latency and reasoning depth.

\paragraph{Extract Method}
Extraction replaces a selected code fragment with a call to a
new helper. Values read by the fragment may become parameters,
and values needed afterward may become return values. If the
helper returns a reference into a borrowed collection, its
signature must preserve the relationship between the reference
and the collection. Extraction also needs to preserve control
flow that crosses the fragment boundary, such as an early return
from the enclosing function~\cite{thy2023adventure}.

\paragraph{Inline Method}
Inlining replaces a call with the callee's computation at the
call site. If an argument has a side effect and the callee uses
the corresponding parameter twice, direct substitution would
evaluate the argument twice. A local binding can preserve the
original single evaluation. Changes to drop scopes can also
change when a value is destroyed, such as when a lock guard
releases its lock~\cite{rust_reference_destructors}. The
\texttt{rust-analyzer} language server provides code actions
for both function extraction and call
inlining~\cite{rust_analyzer_assists}.

\paragraph{Move Method}
Moving a method relocates its implementation and updates how
callers reach it. A cross-module move can change access to
private fields and helpers. A cross-crate move can also require
changing the visibility of the destination method. The resulting
implementation must preserve relevant side effects, return values,
and the contract with its callers.

\subsection{Automated and Verified Refactoring in Rust}

REM repairs control-flow, ownership, and lifetime issues
introduced by method extraction~\cite{thy2023adventure}.
REM-V builds on this work and uses Coq to reason about
observational equivalence for supported
extractions~\cite{britton2025verifying}. REM2.0 extends this
line of work with a \texttt{rust-analyzer}-based extraction
toolchain and an optional equivalence-proof pipeline for a
supported Rust subset~\cite{britton2026rem2}. In our study,
task-specific behavioral tests and structural checks provide
the acceptance criteria for all three refactoring operations.

\subsection{LLM-Assisted Refactoring and Patch Review}

EM-Assist combines model suggestions, static analysis, and
IntelliJ's refactoring engine for Java method extraction.
The model proposes candidate fragments, static analysis
filters and refines the proposals, and the IDE applies the
selected transformation~\cite{pomian2024nextgen}. CodeReviewer
uses pre-training on code changes and reviews to support
change-quality estimation and review-comment
generation~\cite{li2022codereviewer}.

RefAgent coordinates specialized agents for planning, applying,
testing, and refining refactorings. Its evaluation compares
multi-agent and single-agent approaches on open-source Java
projects~\cite{oueslati2026refagent}. Our paired experiment
focuses on the review stage of a Rust refactoring workflow.
Giving each configuration the same patches makes it possible
to measure which faults are caught before execution, which
reference patches are rejected during review, and how review
changes the work performed by the validator.

\section{Approach}
\label{sec:approach}

\begin{figure}[t]
  \centering
  \includegraphics[width=0.92\columnwidth]{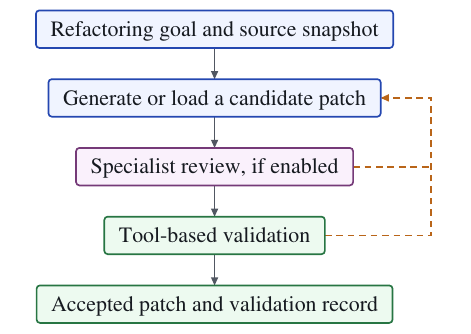}
  \caption{Rust refactoring workflow. The solid arrows show the path
  to acceptance. The dashed arrows show how a review rejection or
  validation failure feeds into a revised candidate when the repair
  budget permits.}
  \label{fig:workflow}
\end{figure}

Our system refactors Rust code given a Cargo project, a
natural-language goal, and a set of editable source files. It
supports method extraction, inlining, and moving through a common
workflow. Figure~\ref{fig:workflow} shows how the orchestrator
obtains a candidate, optionally sends it to specialist review, and
validates it against the original program. A failed candidate can
be revised using evidence from the stage that rejected it. An
accepted candidate is exported as a unified diff together with
its validation record.

\subsection{Task Context and Patch Representation}
\label{sec:candidate}

The orchestrator captures the original contents of the editable
files and provides the relevant source and task goal to the LLM.
For a configured target symbol, it can also include a
Tree-sitter abstract syntax tree and the Rust compiler's mid-level
intermediate representation (MIR). These representations provide
additional context about the structure and control flow of the
target code.

A candidate consists of a list of edits. Each edit specifies a
project-relative path, an original fragment, and a replacement.
Before staging the candidate, the orchestrator checks that each
path belongs to the editable set, that each original fragment
matches exactly one occurrence at the point of application, and
that the source files still match the captured snapshot. It then
applies the edits to a temporary copy of the project and produces
a unified diff. A supplied candidate uses the same representation,
which allows the same patch to be evaluated under different
review policies.

\subsection{Agent Configurations}
\label{sec:agent-configurations}

The configurations use the same underlying model. They differ in
the responsibilities assigned to each model call and in where
those calls occur in the workflow. Specialist roles cover
ownership and lifetimes, concurrency, and performance.

\begin{itemize}
  \item \textbf{Single agent.} A proposer generates the initial
        candidate, which proceeds directly to tool-based
        validation. If a candidate is supplied, the orchestrator
        sends it directly to validation.
  \item \textbf{Designated specialist.} A designated reviewer
        examines the generated or supplied candidate and returns
        an acceptance or rejection decision with its rationale.
        A rejection stops the candidate before execution.
  \item \textbf{Adaptive collaboration.} Routing rules select
        specialists based on the task goal, the changed code, and
        the validation settings. For example, changes involving
        synchronization can trigger concurrency review. The
        selected specialists review the same candidate in sequence.
        The first rejection stops the review. If no rule selects
        a specialist, the candidate proceeds directly to validation.
  \item \textbf{Fixed team.} During end-to-end generation, a
        planner assigns subtasks to specialists, who work from
        the same source snapshot. The orchestrator merges
        non-overlapping edits and requests reconciliation when
        proposed edits overlap. The resulting candidate undergoes
        review and validation. During the paired review experiment,
        the fixed team acts as a review panel for the supplied
        patch, with no planning or patch generation.
\end{itemize}

\subsection{Review Evidence and Scheduling}
\label{sec:review-evidence}

The orchestrator separates three decisions that rely on different
evidence. The first is whether a candidate can be materialized
against the source snapshot. An edit identifies a file and a source
fragment, and replacement is defined only when that fragment
matches uniquely at the point of application. This rule is
important for small edits such as substituting a helper name:
the same name can appear in several call sites in the file. The
second decision is whether a reviewer has identified an issue
worth blocking. The third is whether the staged candidate
satisfies the configured checks. When review is enabled, a
reviewer's approval advances the candidate to the third decision.
A reviewer does not determine the outcome of the downstream test.

Adaptive scheduling uses the goal, the changed fragments, and
the validation settings as routing inputs. Its rules can identify
signals associated with borrowing, visibility, synchronization,
or work on a performance-sensitive path. Where parsing is available,
the router can avoid selecting a specialist merely because a
keyword appeared in a comment or literal. The assignment includes
the reason a reviewer was selected and the constraint to examine.
Review stops at the first rejection. This design can save later
review calls and execution work. It also means that a local
semantic fault can reach the validator if the patch contains no
signal covered by the routing rules.

The two fixed-team uses in this paper occur at different points
in the workflow. For end-to-end generation, a planner assigns
subtasks and the specialists can contribute edits. Their
proposals are merged before validation, with a reconciliation
call when edits overlap. For a supplied patch, the team is a
panel that examines an unchanged candidate in a fixed order.
The paired audit exercises the panel and measures its review
decisions. Model calls made to split or reconcile work belong
to the end-to-end generation experiment.

\subsection{Tool-Based Validation}
\label{sec:tool-validation}

For candidates that proceed to execution, the orchestrator runs
the configured Cargo command on the original and candidate
versions in network-disabled Docker containers. The original run
establishes that the configured check passes on the baseline.
In the two main experiments, the execution check is a
task-specific Cargo test target. The evaluation harness also
checks the diff for the structural changes required by the task.

The system supports additional checks when enabled. Differential
checks compare observations collected from both versions, and a
performance gate can reject measured regressions above a specified
threshold. An optional Z3 check evaluates a predefined property
for supported source patterns. For example, it can check whether
the original and candidate choose the same branch for modeled
lengths and thresholds. In the agent-assisted mode, the model
selects a property from a predefined catalog. The validator
extracts the relevant expressions and invokes the solver.
The solver result applies to the selected property and its
modeled inputs. The main experiments use the task-specific
tests and structural checks described in
Section~\ref{sec:experimental-design}.

\subsection{Repair and Traceability}
\label{sec:validation-repair}

When review or validation rejects a candidate, the orchestrator
can pass the previous proposal and the failure evidence to a model
call for repair. In adaptive mode, the orchestrator can route the
repair to the specialist who raised the issue or to a specialist
associated with the validation failure. If no specialist can be
assigned, the proposer handles the repair.

Each revised candidate is evaluated against the same original
source snapshot. The loop ends when a candidate passes the
configured checks, the repair budget is exhausted, or the
baseline fails. The run record links each revision to the
previous attempt and stores the review decisions and validation
outcomes. These records show where a patch was rejected and what
evidence was available when a revision was requested.
\subsection{Worked Example: Repair After a Failed Check}
\label{sec:worked-repair}

The \texttt{petgraph} reverse-iteration task illustrates how the
system uses validation feedback to revise a candidate. The task
is to inline \texttt{edge\_index} at a single call site in
\texttt{EdgeIndices::next\_back}. The transformation must preserve
the number of elements consumed by each call. It must also preserve
the order of the remaining elements when the caller alternates
between advancing from the front and from the back.

We replayed the task with a supplied candidate that combined the
requested inlining with a change to iterator advancement. As shown
in Fig.~\ref{fig:edge-back-repair}, the candidate correctly replaced
the helper passed to \texttt{map}, and also replaced
\texttt{next\_back()} with \texttt{nth\_back(1)}. The latter skips an
additional element. On a four-edge fixture, the first reverse call
returned index \texttt{2}, while the test expected index \texttt{3}.
The candidate compiled and reached the behavioral assertion.

\begin{figure}[t]
  \centering
  \begin{minipage}{0.97\columnwidth}
  \small\textbf{(a) Original call site}
\begin{lstlisting}[style=rust-example]
self.r.next_back().map(edge_index)
\end{lstlisting}
  \small\textbf{(b) Supplied candidate}
\begin{lstlisting}[style=rust-example]
self.r.nth_back(1).map(EdgeIndex::new)
\end{lstlisting}
  \small\textbf{(c) Revised candidate}
\begin{lstlisting}[style=rust-example]
self.r.next_back().map(EdgeIndex::new)
\end{lstlisting}
  \end{minipage}
  \caption{Feedback-driven repair in the \texttt{petgraph}
  inlining task. The supplied candidate changed the iterator's
  advancement. The revision restored \texttt{next\_back()} while
  retaining the inlined constructor.}
  \label{fig:edge-back-repair}
\end{figure}

The repair call received the previous proposal and the failed
validation output. It restored \texttt{next\_back()} and kept
\texttt{EdgeIndex::new} at the call site. The revised candidate
passed the configured behavioral test, including checks for the
first reverse result, the remaining length, interleaved forward
and reverse iteration, and eventual exhaustion. It also passed
the structural check for the requested inlining.

\begin{figure*}[t]
  \centering
  \includegraphics[width=0.96\textwidth]{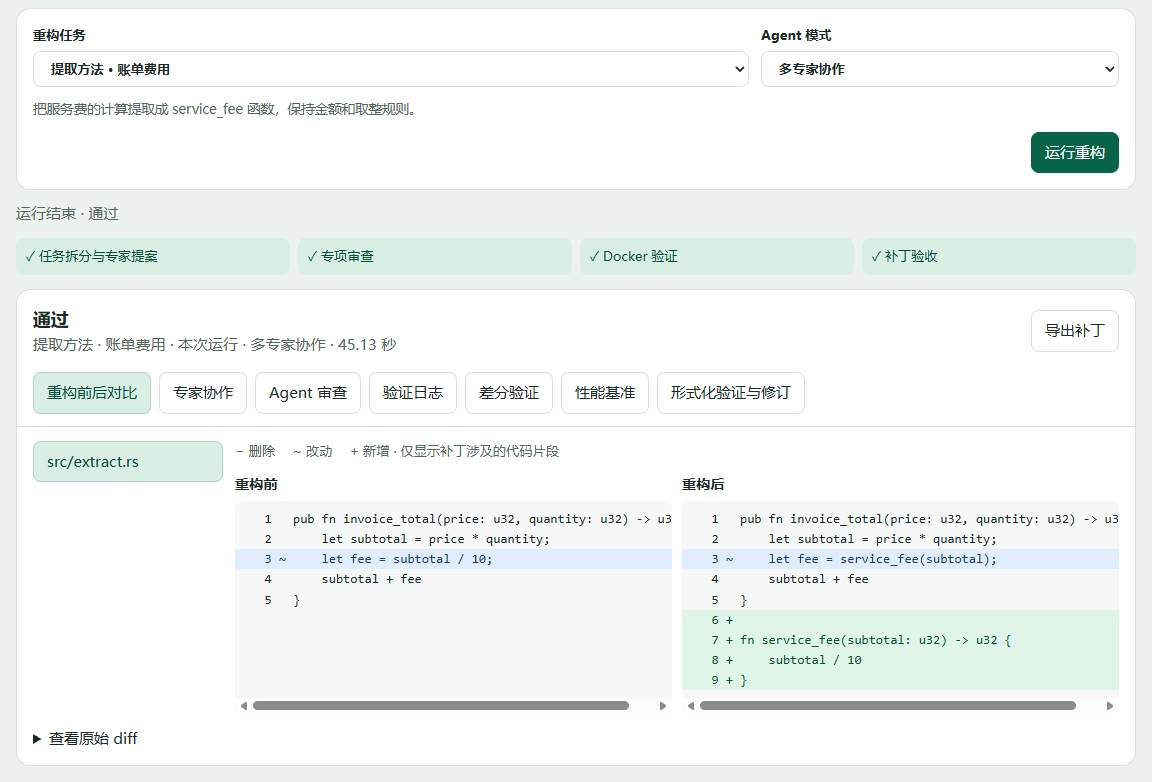}
  \caption{Workbench view of an Extract Method task in multi-agent
mode, showing the execution stages, the side-by-side code diff,
and the accepted patch.}
  \label{fig:workbench}
\end{figure*}

This replay started from a known faulty patch and completed after
one revision. It illustrates the path from an observed failure
to a new patch and a fresh validation run. The end-to-end
generation experiment and the paired review experiment report
their outcomes separately.Figure~\ref{fig:workbench} shows the workbench after an Extract
Method task in multi-agent mode using openPangu-2.1-Pro. The
candidate patch moves the service-fee calculation from
\texttt{invoice\_total} into a private helper, \texttt{service\_fee}.
The interface shows the original and refactored code side by side
and records that the patch passed specialist review and
Docker-based validation.
\section{Experimental Design}
\label{sec:experimental-design}

The two main experiments use the same six Rust refactoring tasks.
The end-to-end experiment compares three complete workflows, from
candidate generation to acceptance. The paired review experiment
supplies the same patches to four policies and uses the same
downstream validation for candidates that proceed to execution.
It measures how review changes the stage of rejection and the
cost of processing a supplied patch.

The worked repair in Section~\ref{sec:worked-repair} traces one
revision prompted by a failed behavioral check. The supplementary
evaluation in Section~\ref{sec:cfg-evaluation} examines a later
workflow revision on four additional tasks. That revision changes
both the generation instructions and the syntax-based preflight
check. The supplementary results therefore describe the combined
revision.

\subsection{Research Questions}

\begin{itemize}
  \item \textbf{RQ1 (End-to-end refactoring).}
        How do the three generation configurations compare in
        observed task acceptance, model calls, and elapsed time
        on the six tasks?
  \item \textbf{RQ2 (Specialist review).}
        Given identical candidate patches, how often does each
        review policy reject a behavior-changing patch before
        execution or stop a reference patch that passes the task
        checks? How do the policies differ in model calls,
        Docker invocations, and elapsed time?
\end{itemize}

\subsection{Tasks and Validation Criteria}
\label{sec:evaluation-tasks}

We select six tasks from fixed snapshots of three open-source Rust
projects, with two tasks for each refactoring operation
(Table~\ref{tab:evaluation-tasks}). The upstream revisions are
\texttt{d6489f248fb0} for \texttt{petgraph},
\texttt{f16da569fcb8} for \texttt{rustc-hash}, and
\texttt{a28f40fdddb7} for \texttt{crossbeam}. The
\texttt{petgraph} extraction task is adapted from the REM
artifact~\cite{thy2023adventure}. The bulk-loop inlining task in
\texttt{rustc-hash} and the two cross-crate move tasks in
\texttt{crossbeam} use fixtures that introduce a helper or wrapper
to be refactored. Each task specifies the requested transformation,
the editable files, and the behavior that must be preserved.

\begin{table}[t]
  \caption{Refactoring tasks and behaviors checked in the evaluation.}
  \label{tab:evaluation-tasks}
  \centering
  \small
  \begin{tabular}{@{}llp{0.50\columnwidth}@{}}
    \hline
    Operation & Project & Behavior checked \\
    \hline
    Extract & \texttt{petgraph}
      & Iterator advancement and access to borrowed edge weights \\
    Extract & \texttt{rustc-hash}
      & Hash output at short-input boundaries \\
    Inline & \texttt{petgraph}
      & Reverse iteration interleaved with forward iteration \\
    Inline & \texttt{rustc-hash}
      & Byte decoding and read order in the bulk loop \\
    Move & \texttt{crossbeam}
      & Backoff state transitions and return values \\
    Move & \texttt{crossbeam}
      & Preservation of the rejected payload and queue contents
        when the queue is full \\
    \hline
  \end{tabular}
\end{table}

For each task, we prepare a reference patch and derive a
behavior-changing variant by modifying that patch. Before
evaluation, we verify that the original program and reference
patch pass the same task-specific Cargo test target. We also
verify that the behavior-changing variant compiles and fails
a behavioral assertion in that target. Thus, every faulty
variant in the paired experiment has a known downstream test
that can detect it.

The evaluation harness uses task-specific checks on the diff to
confirm that an accepted patch performs the requested structural
change. For example, an extraction must introduce the requested
helper and call it from the original method. Acceptance in
these experiments is defined by the task's behavioral and
structural checks.
\subsection{Construction of the Paired Patches}
\label{sec:paired-patch-construction}

We construct each behavior-changing candidate by modifying the
reference patch for the same task. Both candidates therefore
contain the requested kind of refactoring and target the same
editable files. The additional modification changes an operation
within that refactoring, such as the index passed to an iterator,
a boundary comparison, or the action taken when a queue is full.
This pairing allows reviewers to encounter a structurally
plausible patch whose fault depends on the semantics of the change.

Table~\ref{tab:seeded-faults} links each modification to the evidence
used to qualify it. The tests observe return values and state
visible to callers, including iterator consumption and queue
contents. For the borrowed-edge extraction, the test also checks
that the returned weight refers to the original graph entry.
For the hashing tasks, the tests compare the resulting hash for
selected inputs against an independently computed expectation.

\begin{table*}[t]
  \caption{Construction and qualification of the six seeded faults.
  Each mutant is derived from the reference patch for the same task.}
  \label{tab:seeded-faults}
  \centering
  \small
  \begin{tabular}{@{}p{0.20\textwidth}p{0.30\textwidth}
                    p{0.43\textwidth}@{}}
    \hline
    Task & Seeded change & Evidence from the behavioral test \\
    \hline
    Borrowed-edge extraction
      & Pass \(n+1\) to the inner iterator for the tested index.
      & The call returns edge \texttt{(3, 4)} instead of the
        expected \texttt{(2, 3)}. The test also checks reference
        identity and subsequent iterator operations. \\
    Short-input hash extraction
      & Change the 8-byte branch condition from
        \texttt{len >= 8} to \texttt{len > 8}.
      & The hash is compared with the expected result for
        every input length from 0 through 16. \\
    Reverse-edge inlining
      & Use \texttt{nth\_back(1)} for the reverse advance.
      & The first call returns index 2 where index 3 is
        expected. The test also checks length and the order
        of interleaved advances. \\
    Bulk hash inlining
      & Decode the first word a second time when computing
        the value for the second word.
      & Bulk-input hashes are checked against expectations
        computed from words read in source order. \\
    Backoff method move
      & Return \texttt{false} after taking a pending backoff step.
      & The test compares the wrapper's return value and
        completion state with the underlying backoff behavior. \\
    Queue method move
      & Replace rejection on a full queue with an overwrite
        and report success.
      & The test expects the original rejected payload and
        checks that existing items remain in queue order. \\
    \hline
  \end{tabular}
\end{table*}

This qualification procedure establishes a known outcome for
each paired patch. Passing both the original and reference
versions confirms that the test accepts the intended behavior
before and after the transformation. Requiring the mutant to
compile and fail a behavioral assertion excludes failures
attributable only to syntax or type errors. The resulting set
supports an analysis of when a known fault is detected. The
coverage of faults in naturally generated patches remains a
separate empirical question.
\subsection{End-to-End Refactoring (RQ1)}

We run the single-agent, adaptive, and fixed-team configurations
once on each of the six tasks, yielding 18 runs. All three
configurations receive the same task goal and editable source
snapshot for a given task. Each has a budget of one repair
attempt after the initial proposal. The fixed team participates
in candidate generation in this experiment.

We record first-attempt acceptance, final acceptance after any
repair, the total number of model calls, and end-to-end
wall-clock time. This experiment compares complete workflows,
including any differences in both candidate generation and
review.

\subsection{Paired Patch Review (RQ2)}

We give each configuration the same six reference patches and
six behavior-changing variants, yielding 48 runs. The four
configurations use direct validation, one designated specialist,
adaptive specialist selection, or a fixed specialist panel.
The fixed panel reviews the supplied patch without running
the planning or generation stages. The harness uses patch hashes
to confirm identical candidate inputs, keeps the reference
labels outside the reviewer input, and rotates configuration
order across patches. Repair is disabled so that each outcome
reflects the disposition of the supplied patch.

We count an \emph{early rejection} when review stops a
behavior-changing patch before Docker validation. We count a
\emph{false rejection under the behavioral reference label}
when review stops a reference patch that satisfied the task
checks. We also record final disposition, model calls, Docker
invocations, and end-to-end wall-clock time. Each execution of
a baseline or candidate command in Docker counts as one
invocation. Elapsed time includes review, staging, and
execution for the stages reached by the candidate.

Because the behavior-changing variants are qualified using the
same downstream tests, the paired comparison measures how
review changes the stage of rejection and its cost. The six
reference patches and six faulty variants form a deliberately
balanced workload for comparing the policies.

\subsection{Execution Environment and Artifacts}

The workflow uses a configurable LLM backend for candidate generation
and specialist review. The current workbench defaults to
openPangu-2.1-Pro, accessed through Huawei Cloud ModelArts MaaS
using the model identifier \texttt{openpangu-2.1-pro}. The configured Cargo test targets run
offline in local Linux containers using the \texttt{rust:1.85}
image. The main experiments use task-specific tests and
structural checks. The optional Z3 and performance gates are
outside the scope of these experiments.

The \texttt{evaluation/dataset-pilot/} directory contains the
task definitions, fixtures, and behavioral tests. Separate JSON
records in \texttt{reports/} document task qualification,
end-to-end refactoring, and paired patch review. The paired
review records include the input patch hashes and per-run review
and validation outcomes.

\subsection{Pairing and Outcome Interpretation}
\label{sec:pairing}

The unit of comparison for specialist review is a supplied
patch under one review policy. For each task, the reference
patch implements the requested restructuring. A narrow change
to its implementation produces the faulty variant while
retaining the visible structure of the refactoring. For example,
the extraction still introduces the requested helper after the
iterator error is inserted. A reviewer therefore cannot infer
behavioral correctness merely by checking whether the expected
helper or call site is present. The qualification results
establish that the downstream test passed on the original and
reference programs and failed on the faulty variant. This
selection procedure makes the review-stage comparison
interpretable and conditions it on faults already detectable
by the task tests.

The review harness retains the patch labels for scoring and gives
each policy the same patch content for a given input. Each
candidate that reaches execution is subject to the same task
test. The audit records a reference patch stopped at review
as a false rejection under the behavioral label. This label
establishes that the patch passed the available task checks.
It does not resolve questions about behavior outside those
checks, including performance. Model-call count measures the
number of requests rather than token consumption. A Docker
invocation is one recorded baseline or candidate command,
whereas wall-clock time captures the complete path taken by a
candidate.

\subsection{Structural Constraints and Behavioral Checks}

The structural checks serve a different purpose from the test
target. A patch that leaves the original method in place and
adds an unused helper could preserve every tested observation.
It would still fail the extraction task. Conversely, a patch
can introduce the required helper and replace the original
call while changing the value returned to callers. The paired
faulty variants are examples of the latter case. The evaluation
therefore checks both the requested transformation and the
observations that transformation must preserve.

\begin{table}[t]
\caption{Examples of acceptance evidence by refactoring type.
The structural checks inspect the candidate diff.}
\label{tab:criteria-examples}
\centering
\small
\begin{tabular}{@{}lp{0.36\columnwidth}p{0.36\columnwidth}@{}}
\hline
Type & Structural evidence & Behavioral evidence \\
\hline
Extract & Introduce the helper and call it from the original
method. & Preserve which edge is returned and keep its weight
borrowed from the graph. \\
Inline & Remove the selected mapping through the old helper
and construct the edge index at the site. & Return the last
remaining edge and preserve the subsequent iteration order. \\
Move & Remove the old wrapper method and delegate to the
relocated implementation. & Return the rejected payload on a
full queue and preserve queued elements. \\
\hline
\end{tabular}
\end{table}

Table~\ref{tab:criteria-examples} illustrates how the two kinds
of evidence complement one another. Structural acceptance is
task-specific. In the reverse-iterator example, the refactoring
applies to one call site, and other callers can continue using
the existing helper. In a move, a search for the old function
name alone is insufficient because the destination may keep the
same name. Checks can instead distinguish removal at the old
location, introduction at the new location, and delegation
through the unchanged public entry point.

The behavioral inputs are designed to distinguish the behavior
that the injected mistake would change. In the borrowed-edge
task, distinct endpoints expose an extra advancement. A pointer
comparison checks that the returned edge weight is the value
stored in the graph. In the full-queue task, distinct payloads
let the test distinguish a rejected insertion from an overwrite.
These assertions support the conclusions about the selected
inputs and fault mechanisms. They do not characterize every
operation that the original or candidate can perform.

\subsection{Supplementary Conditional-Compilation Evaluation}
\label{sec:cfg-evaluation}

We also examine four additional tasks from the same
\texttt{rustc-hash} and \texttt{crossbeam} revisions. These
tasks exercise two inlining operations and two moves. One
inlining operation replaces four helper calls in
\texttt{FxHasher::write\_u128}, preserving the order of the hash
updates and the separate pointer-width guards. The other
inlines the queue-capacity query at a single site in
\texttt{ArrayQueue}'s consuming iterator. The two moves
relocate the public \texttt{FxHasher::with\_seed} constructor
and the \texttt{Default} implementation for
\texttt{SegQueue}, respectively. The former preserves its
public \texttt{const} signature, and the latter preserves its
allocation and atomic-pointer configuration conditions.

We compare two successive workflow versions on these four tasks.
Each version has two runs per task and generation configuration,
for 24 runs per version. These follow-up runs use
\texttt{openpangu-2.1-pro}, a budget of one repair, offline Cargo
validation, and common task-specific structural acceptance rules.
The revised version includes a generation instruction about
preserving statement-level configuration attributes and a
syntax-based preflight check that supplies targeted diagnostics
when that constraint is violated. The comparison treats these
two changes as a single workflow revision. The experiment also
includes a separate 32-bit check of the original program and
the repaired \texttt{write\_u128} example. The goal of that
check is to exercise a target-specific branch that the
ordinary host-target runs do not execute.

\section{Results}
\label{sec:results}

\subsection{RQ1: End-to-End Refactoring}

All three configurations passed all six tasks on their first
attempt (Table~\ref{tab:generation-results}). Each configuration
passed both extraction tasks, both inlining tasks, and both move
tasks. None of the 18 runs used the available repair attempt.

\begin{table}[t]
  \caption{End-to-end results over six tasks per configuration.
  Model calls are summed across tasks.}
  \label{tab:generation-results}
  \centering
  \small
  \begin{tabular}{@{}lrrr@{}}
    \hline
    Configuration & First pass & Calls & Median (s) \\
    \hline
    Single agent & 6/6 & 6  & 39.9 \\
    Adaptive     & 6/6 & 13 & 58.5 \\
    Fixed team   & 6/6 & 31 & 71.0 \\
    \hline
  \end{tabular}
\end{table}

The single-agent configuration made six model calls across the
six tasks, compared with 13 for adaptive collaboration and 31
for the fixed team. It also had the lowest median end-to-end
time at 39.9\,s. Adaptive collaboration took 58.5\,s and the
fixed team took 71.0\,s at the median. The three workflows
therefore differed in observed cost while obtaining the same
acceptance count under the task checks.

\subsection{RQ2: Paired Patch Review}

The designated specialist and fixed panel each rejected six of
six behavior-changing patches before Docker validation
(Table~\ref{tab:review-results}). Adaptive collaboration rejected
five at review. The remaining patch changed reverse iteration
in \texttt{petgraph}, and the task-specific test caught the fault.
In the direct-validation configuration, the tests caught all six
faults, as expected from the task qualification procedure.

\begin{table*}[t]
  \caption{Paired patch-review results. Each configuration received
  six reference patches and six behavior-changing variants. False
  rejections are counted under the behavioral reference labels.
  Model calls, Docker invocations, and wall-clock time are summed
  across the 12 runs per configuration.}
  \label{tab:review-results}
  \centering
  \small
  \begin{tabular}{@{}lrrrrr@{}}
    \hline
    Configuration & Early rejections & False rejections &
    Model calls & Docker invocations & Total time (s) \\
    \hline
    Single agent (direct validation) & 0/6 & 0/6 &  0 & 24 & 148.9 \\
    Designated specialist            & 6/6 & 0/6 & 12 & 12 & 129.5 \\
    Adaptive                         & 5/6 & 0/6 & 12 & 14 & 163.1 \\
    Fixed panel                      & 6/6 & 2/6 & 24 &  8 & 225.0 \\
    \hline
  \end{tabular}
\end{table*}

The designated specialist and adaptive collaboration allowed all
six reference patches to proceed. The fixed panel rejected two
reference patches that had passed the task checks. In both cases,
the performance specialist raised concerns about compiler inlining
of a newly introduced or relocated helper. One patch extracted
short-input hashing logic in \texttt{rustc-hash}. The other moved
a backoff method across crates in \texttt{crossbeam}. The
experiment did not benchmark those potential regressions, so the
reported false rejections are relative to the behavioral
reference labels.

On the balanced set of 12 patches, the designated specialist
reduced Docker invocations from 24 to 12 and recorded the lowest
total wall-clock time. Relative to direct validation, it added
5.6\,s on average for a reference patch and saved 8.8\,s on
average for a behavior-changing patch. The fixed panel made the
fewest Docker invocations and took the longest overall. Its
invocation count reflects both the six early fault rejections
and the two reference patches it stopped before execution.

Figure~\ref{fig:petgraph-review-example} shows the
\texttt{petgraph} extraction task. The reference patch moves the
selected code into a private helper and preserves the lifetime of
the returned edge weight. The seeded variant changes the helper's
iterator advance. The designated specialist identified that change
and rejected the patch in 2.5\,s without starting Docker validation.
Direct validation rejected the same patch in 13.5\,s when the
behavioral assertion failed. Both times are end-to-end measurements
for these two runs.

\begin{figure*}[t]
  \centering

  \begin{minipage}[t]{0.48\textwidth}
    \textbf{(a) Before extraction}\par\smallskip
\begin{lstlisting}[style=rust-example]
fn nth(&mut self, n: usize)
    -> Option<Self::Item> {
    // Selected fragment
    self.inner.nth(n).map(
        |(&(n1, n2), weight)| (n1, n2, weight)
    )
}
\end{lstlisting}
  \end{minipage}\hfill
  \begin{minipage}[t]{0.48\textwidth}
    \textbf{(b) Reference extraction}\par\smallskip
\begin{lstlisting}[style=rust-example]
fn nth(&mut self, n: usize)
    -> Option<Self::Item> {
    self.hard_nth_edge(n)
}

fn hard_nth_edge(&mut self, n: usize)
    -> Option<(N, N, &'a E)> {
    self.inner.nth(n).map(
        |(&(a, b), weight)| (a, b, weight)
    )
}
\end{lstlisting}
  \end{minipage}

  \par\smallskip
  \begin{minipage}{0.98\textwidth}
    \small
    \textbf{Seeded fault in (b).}
    Replace \texttt{.nth(n)} with
    \texttt{.nth(n.saturating\_add(1))}.
    For \texttt{nth(1)}, the test expects edge
    \texttt{(2, 3)}, while the modified helper returns
    \texttt{(3, 4)}.
  \end{minipage}

  \caption{Extract Method example from
  \texttt{petgraph}'s \texttt{AllEdges::nth}. The listings
  show the relevant methods and omit their enclosing
  \texttt{impl} blocks.}
  \label{fig:petgraph-review-example}
\end{figure*}
\subsection{Review Outcomes by TaskReview Outcomes by Task}
\label{sec:review-detail}

Table~\ref{tab:review-matrix} shows the rejection stage for
each behavior-changing variant. The designated specialist and
the fixed panel caught all six variants at review. Adaptive
collaboration caught five. Its routing rules selected no
specialist for the reverse-iteration patch, so the candidate
proceeded directly to the task test.

The missed patch replaced \texttt{next\_back()} with
\texttt{nth\_back(1)}, which consumes an additional element
from the back of the iterator. The first assertion expected
edge index 3 and received edge index 2. The designated
specialist and the fixed panel identified the same fault
before execution. In this case, the difference arose at the
routing stage: the patch reached validation without a
specialist assessment.

\begin{table}[t]
\caption{Rejection stage by task for faulty variants.}
\label{tab:review-matrix}
\centering
\small
\setlength{\tabcolsep}{4pt}
\begin{tabular}{@{}lcccc@{}}
\hline
Task & Direct & Specialist & Adaptive & Panel \\
\hline
Borrowed edge & T & R & R & R \\
Short-input hash & T & R & R & R \\
Reverse iterator & T & R & T & R \\
Bulk hash inlining & T & R & R & R \\
Backoff method move & T & R & R & R \\
Queue method move & T & R & R & R \\
\hline
\end{tabular}

\smallskip
\raggedright\footnotesize R: rejected at review. T: rejected by the task test.
\end{table}

The two disagreements on reference patches occurred in the
fixed panel. For the short-input hash extraction, its performance
reviewer raised the possibility that the new helper would remain
out of line in downstream builds. For the backoff move, the
reviewer raised a similar concern about the cost of a new
cross-crate call. Both references had satisfied the behavioral
checks before the audit. The other policies allowed them to
reach validation. The panel decisions therefore illustrate a
difference in the evidence used to block a candidate. A
specific test can demonstrate that an iterator consumed the
wrong item on an input. An argument about inlining identifies a
measurement to make and leaves the existence and magnitude of
the runtime effect open.

The balanced workload also explains the difference between
execution effort and elapsed time. The designated specialist
removed twelve Docker invocations by stopping the six faulty
patches early. The fixed panel removed sixteen invocations:
twelve from faulty patches and four from the two stopped
references. The latter reduction comes with lower acceptance
of the reference patches. In addition, the recorded review
stage took 62.0\,s for the designated specialist, 82.6\,s
for adaptive collaboration, and 176.7\,s for the fixed panel
across their 12 runs. The totals include the variability of
individual model calls. The fixed panel thus did less
container work and still took more wall-clock time than
direct validation.

\subsection{Timing by Patch Pair}

The aggregate results combine different execution costs and
review paths. Table~\ref{tab:per-patch-time} reports both
members of each pair so that the dependence on workload is
visible. For example, on the seeded short-input hash fault,
the direct path took 6.0\,s and the specialist path took
4.5\,s. The corresponding reference took 5.9\,s directly
and 19.0\,s with the specialist. On the queue fault, the
direct path took 20.1\,s and the specialist path took
3.7\,s. The same review policy therefore had different
timing consequences across tasks, even when its decision
agreed with the task labels in every case.

\begin{table*}[t]
\centering
\caption{End-to-end wall-clock time by supplied candidate.
Each cell gives reference and faulty-variant times, respectively,
in seconds. Bold reference times denote rejections at review.}
\label{tab:per-patch-time}
\small
\begin{tabular}{@{}lrrrr@{}}
\hline
Task & Direct validation & Designated specialist & Adaptive & Fixed panel \\
\hline
Borrowed-edge extraction & 18.0 / 13.5 & 18.2 / 2.5 & 20.4 / 3.3 & 33.7 / 8.6 \\
Short-input hash extraction & 5.9 / 6.0 & 19.0 / 4.5 & 24.7 / 5.2 & \textbf{38.4} / 7.0 \\
Reverse iterator inlining & 12.3 / 12.0 & 15.4 / 2.8 & 12.2 / 13.3 & 28.6 / 3.5 \\
Bulk hash inlining & 5.7 / 5.8 & 14.1 / 3.6 & 24.4 / 4.1 & 31.5 / 2.7 \\
Backoff method move & 20.8 / 14.8 & 22.0 / 2.3 & 25.7 / 3.7 & \textbf{22.9} / 2.7 \\
Queue method move & 14.0 / 20.1 & 21.5 / 3.7 & 21.8 / 4.3 & 39.0 / 6.5 \\
\hline
\end{tabular}
\end{table*}

The review explanations also help interpret what was caught.
For the short-input hash helper, the designated specialist
identified the changed boundary from \texttt{>= 8} to
\texttt{> 8}. At length eight the faulty helper follows the
four-byte path instead of the eight-byte path. In the bulk
hash case, the cursor advances twice, yet the second decoding
uses the slice bound by the first read. The reviewer
identified the incorrect operand even though the number of
reads and the cursor movement still resemble the original
structure. These two faults require different checks:
boundary coverage in the first case and distinguishable
input words in the second.

The cross-crate move tasks demonstrate why checking that a
caller delegates to the new location is only part of the
acceptance criterion. The faulty backoff method delegates to
the relocated code but returns the wrong Boolean after
advancing the state. The faulty queue method delegates to
an operation that can overwrite the oldest item on a full
queue and reports success. The designated specialist
identified both contract violations. The tests exercise
the corresponding paths by comparing the wrapper's return
value and completion state, and by checking that a full
queue returns the submitted payload and retains the elements
it already holds. These checks locate effects at callers of
the moved methods, where the behavioral consequences of the
edits are observable.

\subsection{Supplementary Results on Conditional Compilation}
\label{sec:cfg-results}

Table~\ref{tab:cfg-followup} reports the two workflow versions
from Section~\ref{sec:cfg-evaluation}. Under the original
workflow, 18 of 24 runs passed on the first attempt and the
other six passed after one repair. Every run that needed a
repair was on \texttt{write\_u128}, with two such runs per
configuration. In the revised workflow, all 24 runs passed on
their first attempt. The other three tasks passed on the
first attempt in all 18 runs per version. Thus the observed
difference was concentrated in the task that exercises
statement-level pointer-width conditions.

\begin{table}[t]
\caption{Supplementary runs before and after the workflow revision.
Each configuration ran eight times per version.}
\label{tab:cfg-followup}
\centering
\small
\setlength{\tabcolsep}{4pt}
\begin{tabular}{@{}lccc@{}}
\hline
Configuration & First pass & Calls & Median (s) \\
\hline
Single agent & $6\to8$ & $10\to8$ & $25.9\to24.3$ \\
Adaptive & $6\to8$ & $10\to8$ & $29.3\to24.4$ \\
Fixed team & $6\to8$ & $44\to40$ & $59.4\to52.1$ \\
\hline
\end{tabular}
\end{table}

The \texttt{write\_u128} task illustrates why the additional
check is useful. A recorded faulty candidate attaches a
\texttt{cfg} attribute directly to an assignment. The stable
compiler rejects that form in the edited context. Feeding back
the diagnostic allowed the model to wrap the conditional
assignments in guarded blocks and retain the ordering of the
original state updates. A separate check on the
\texttt{i686-unknown-linux-gnu} target ran the original and
repaired versions. Both passed the package tests and a focused
assertion that uses distinct 32-bit words to check update
order. This check adds execution evidence for the repair on one
32-bit target. The ordinary runs compare the workflows on the
host target and use structural checks for the presence of
the guards.

The first-pass counts in this follow-up are descriptive. The
generation instruction and preflight check were changed
together, and the experiments were run in successive batches.
The comparison therefore characterizes the observed behavior
of the revised workflow as a whole. The earlier paired audit
addresses a different question and retains its original
results and denominators.

\section{Discussion and Threats to Validity}
\label{sec:discussion}

\subsection{Cost Trade-offs of Specialist Review}

All three end-to-end workflows passed the six tasks on their first
attempt. Direct validation required the fewest model calls and had
the lowest median elapsed time in those runs. The paired experiment
shows where specialist review can help: for a candidate with a
detectable fault, a reviewer can stop the patch before staging and
container execution. For a reference patch that proceeds to
validation, review adds a model call while the downstream checks
still run.

The balance depends on the incoming patches. Relative to direct
validation, the designated specialist added 5.6\,s on average per
reference patch and saved 8.8\,s on average per behavior-changing
patch. Let \(p\) be the fraction of incoming patches with faults
comparable to the seeded faults. If the measured average costs
hold, the estimated difference in time per patch, in seconds, is
\[
  \Delta T(p) = 5.6(1-p) - 8.8p.
\]
A negative value favors specialist review on elapsed time. The
resulting break-even fraction is approximately \(39\%\). This is an
illustrative estimate for the observed workload. Changes in model
latency, test runtime, or fault composition can shift the
break-even point.

\subsection{Evidence Requirements for Review Decisions}

The paired audit illustrates how the basis for a rejection
affects its interpretation. In the reverse-iteration case, the
reviewer identified an extra iterator advance and the task test
exhibited its effect on the returned edge. In the two fixed-panel
rejections of reference patches, the reviewer anticipated a
possible runtime cost from changes to compiler inlining. The
patches had passed their behavioral checks, and the audit did
not measure the suspected performance regressions.

For a review policy, these cases suggest different next steps.
A concrete discrepancy against the task's behavioral requirements
can justify early rejection. A performance concern can instead
identify the code path and build configuration for a targeted
benchmark. The benchmark result can then inform the acceptance
decision. Whether this policy reduces unnecessary rejections
remains a question for future evaluation.

\subsection{Implications for Validation Design}

The evaluation exhibits two different uses of feedback. The
reverse-iteration example uses a task-level assertion after a
well-formed Rust expression passes compilation. The
conditional-compilation example uses a compiler diagnostic or
a structural preflight finding about the form of the edit. The
evidence needed to revise the patch is different in each
case. For the former, the assertion identifies the observable
effect of consuming an extra element. For the latter, the
check identifies how to express conditional code while preserving
the target-specific branches. A useful repair stage can retain
that distinction in the feedback it sends to the model and
then re-evaluate the resulting patch under the same task
criteria.

These cases also illustrate how to choose between a structural
check and a specialist decision. A recurring transformation
constraint such as preserving a statement-level configuration
guard can be encoded and validated against the patch. A
performance concern about a newly introduced call depends on
the build and runtime environment and can be tested separately
with a focused benchmark. For a semantic requirement such as
reverse-iteration order, a task-specific assertion remains a
direct source of evidence. The model's review serves as an
earlier decision point over these obligations.

\subsection{Evidence Needed for Generalization}

The study distinguishes three questions for subsequent
evaluation. The first concerns repeatability of a review
decision. It can be investigated by repeating the same
patch--policy pairs and recording variation in rejection
stage, model calls, and elapsed time. Because a single
candidate is shared across policies, a repeated design can
retain the pairing when comparing both binary decisions and
costs. The second concerns fault coverage. A study using
naturally occurring refactoring patches would need independent
behavioral labels, including an explicit category for cases
where tests and reviewer evidence do not resolve the
outcome. Such a set would reveal how review behaves on a
broader range of faults than the deliberately qualified set
used here. The third concerns the costs of accepting or
rejecting an ambiguous case. The unmeasured performance
concerns in the fixed panel provide concrete candidates for
that investigation.

The cost calculation above treats elapsed time as the outcome
and uses averages from the sampled patches. In practice, an
early rejection can transfer work to a subsequent repair or a
human maintainer. A deployment evaluation should therefore
record the work required between an initial review decision
and a candidate that is ready for adoption. This study
isolates the first decision in the paired audit by setting
the repair budget to zero. The generation experiment and
worked repair example provide separate observations of
repair behavior. Together these settings explain why
first-pass review latency and complete repair cost require
separate measurements.

\subsection{Threats to Validity}

\noindent \paragraph{Task and fault selection}
The six tasks were selected to exercise specific obligations in
three Rust projects, and some tasks use fixtures added to the
upstream snapshots. The paired faults were deliberately seeded
and qualified with the downstream tests. This design supports a
controlled comparison of when a known fault is rejected.
Naturally occurring patches may introduce different faults, and
some of those faults may escape the available tests. The balanced
patch set also does not estimate the prevalence of faulty patches
in production.

\paragraph{Acceptance criteria}
Task acceptance is based on behavioral tests and structural
checks. A passing patch satisfies the checks used for its task.
The reported false rejections are therefore relative to the
behavioral reference labels. The main experiments do not measure
whole-program semantic equivalence or runtime performance.

\paragraph{Repeatability and cost}
In the main experiments, we ran each task--configuration and
patch--configuration pair once with one model identifier. The observed counts and timings
do not quantify variation across repeated model calls or different
models. Elapsed time also depends on API latency, build caches,
and the local execution environment. The paired comparison
rotated configuration order across patches, while the timing
results remain specific to the recorded runs. The supplementary
comparison repeats generation on four tasks and varies two workflow
components together. It provides no repeated observations of the
fixed-patch review decisions.
\section{Conclusion}
\label{sec:conclusion}

We presented a LLM-based workflow for extracting, inlining,
and moving methods in Rust. To measure the contribution of
specialist review, we gave four configurations the same candidate
patches and used the same downstream validation procedure for
patches that proceeded to execution.

In the end-to-end experiment, all three generation workflows
passed the six tasks on their first attempt. The single-agent
workflow used the fewest model calls. In the paired review
experiment, a designated specialist rejected all six
behavior-changing patches before Docker validation, allowed all
six reference patches to proceed, and reduced Docker invocations
from 24 to 12 relative to direct validation. The fixed panel
reduced invocations further, to eight, while recording the
highest total elapsed time and rejecting two reference patches
over unmeasured performance concerns.

On the tasks studied, specialist review changed when known
faults were caught and how much work reached execution-based
validation. Its practical value depends on how often candidates
contain faults, how expensive downstream checks are, and how
often review stops patches that satisfy those checks. Future work
should repeat the comparison with naturally occurring patches
and use targeted benchmarks to evaluate performance concerns
raised during review.

\bibliographystyle{IEEEtran}
\bibliography{references}

\end{document}